\documentstyle[psfig]{article}
\begin{document}
\title{ Signatures of dual scaling regimes in a simple avalanche model for
magnetospheric activity.
}
\author{S. C. Chapman$^1$, N. W. Watkins$^2$ and G. Rowlands$^1$\\
$^1$Department of Physics, Warwick University, Coventry CV4 7AL, UK\\
$^2$British Antarctic Survey, Cambridge, CB3 0ET, UK}

\date{\today}
\maketitle
\begin{abstract}
Recently, the paradigm that  the dynamic magnetosphere 
 displays sandpile-type phenomenology has been advanced,
 in which 
energy
dissipation is by means of avalanches which do not have an intrinsic
scale. This may in turn imply that the system is in a
self organised critical (SOC) state. Indicators of internal
processes are consistent with this, examples are the power law
dependence of the power spectrum of auroral indices, and in-situ magnetic field
observations in the earth's geotail. 
 An apparent paradox is that, rather than power laws,
 substorm statistics exhibit probability distributions with
 characteristic scales. Here we discuss a simple sandpile
 model which yields  for energy
 discharges due to internal reorganization a probability distribution
 that is a power law, 
 whereas systemwide discharges (flow of ``sand" out of the system) form a
 distinct
 group whose probability distribution has a well defined mean.
  We analyse the model over a wide   dynamic range whereupon two regimes
 having different inverse power law statistics emerge, corresponding
 to reconfigurations over two distinct scaling regions:  short scale sizes
 sensitive to the discrete
 nature of the sandpile model, and long scale sized
 up to the system size which  correspond to the continuous limit of the
 model. The latter
 are anticipated to correspond to large scale systems such as the magnetosphere.
  Since the energy inflow may be highly
  variable, we examined the response of the model under strong
  or variable loading and established that the power law
  signature of the large scale internal events persists. The interval
  distribution of these events is also discussed.

\end{abstract}

\section{Introduction}

Recently there has been interest in relating observed
characteristics of global energy
transport in space 
plasmas to ``sandpile" models which dissipate energy
by means of avalanches  (Consolini, 1997; Chapman {\em et al.}, 1999).
When such models exhibit scale free, inverse power law
statistics in the probability distributions of energy released by
avalanches, and of avalanche length and duration,
they are candidates for description in terms of 
self organised criticality (SOC) (Bak {\em et al.}, 1987, 1988; Lu, 1995)
see also Jensen (1998) and references therein.  The power
spectra may also have  an inverse power law (``1/f") signature,
and SOC was introduced to explain the ubiquity of such spectra 
and of fractality in Nature.

Chang's suggestion (Chang, 1992, 1998a, 1998b)
that the magnetosphere is in an SOC state
 has motivated application of avalanche
 models to the solar wind-magnetosphere-ionosphere system.
 Observational motivation  includes sporadic
nature of energy release events within the magnetotail (``Bursty Bulk
Flow" events (Angelopoulos {\em et al.}, 1996)),
 power law in-situ magnetic field  power spectra (Hoshino {\em et al.}, 1994),
and power law features of
magnetospheric index data,
notably AE which is an indicator of
 energy dissipated by the
magnetosphere into the ionosphere.
Tsurutani {\it et. al.} (1990)   described a broken power law AE
spectrum;   indicative of SOC but not conclusive as power law power
spectra are not unique to SOC systems (Jensen, 1998) . Consolini (1999)
  has recently used AE
data taken over a ten year period
 to construct the distribution $D(s)$ of a burst
 measure $s$   extending
the result obtained for one year in  (Consolini, 1997). This
work strongly suggests
that inverse
 power law burst statistics are a robust feature of the AE data, albeit with an exponential
 tail and some evidence of an additional lognormal component.
It is currently less clear that the power law is solely of intrinsic
magnetospheric origin and
it may in fact be related to the behaviour of the solar wind energy input 
(Freeman {\em et al.}, 1999).
 Since  the
global reconfigurations of the magnetotail (substorm events)
appear to have occurrence statistics with a well
defined mean,  Chapman {\em et al.} (1998) demonstrated a  
simple avalanche model (another example is  (Pinho and Andrade, 1998)) 
that in principle
has relevance for the magnetosphere in that it   yields systemwide
avalanches where the statistics have a well defined mean (intrinsic scale),
whereas their internal avalanche statistics are scale free.

An additional consideration for space plasmas are
 the means by which the
conjecture of scale free statistics  can be tested. We wish to test
the hypothesis that 
 the probability distributions of energy dissipated, length
scales and duration of avalanches are  of power law 
form as  seen in the original observations of
SOC (Jensen, 1998) in a slowly driven sandpile.
Since to test for power law statistics
we need to maximise the range of sizes of
events, the required  statistical experimental evidence  
requires long runs of data. In the
magnetospheric system   this will then 
imply that both the instantaneous
value and  smoothed local mean of the loading rate
(the solar wind) will have strong variation.

In this paper we  illustrate some  robust features of the avalanche
statistics of the simple avalanche model   (Chapman {\em et al.}, 1998, 1999)
which are needed  for application of such a model to space  
plasma data.  
We investigate to what  extent  the model gives 
inverse power law avalanche statistics   under  
slow loading. 
We show   how these statistics are modified under strong and/or
variable loading. We shall also see that two distinct regimes for energy
transport, both with power law avalanche statistics
but with different slopes, emerge from the
sandpile algorithm, depending on the size of the system.
In addition we show how these may
be characterized in terms of the interval distribution of events
on different spatial scales.

\section{A two-regime avalanche model}
\subsection{The algorithm}

 Sandpile algorithms  generally include
an array of nodes, at each of which there is a variable amount
(height) of sand; a critical gradient (difference in height between
neighbouring nodes) which, if exceeded by the actual gradient, triggers
local redistribution of sand; and algorithms for redistribution and fuelling.
 The main measured output is the statistics
 of the emergent avalanche distribution.
 Kadanoff {\em et al.} (1989) gave 
 an early classification of such models. The relationship 
 of the models  to experimental
sandpiles, and to the ideal concept of SOC remain topics of active research
(see for example  Jensen (1998), also Dendy and Helander (1997)). 

The sandpile  model used here is described in
more detail in  (Chapman {\em et al.}, 1999; Helander {\em et al}, (1999)).
 We have a one-dimensional grid of $N$ equally
spaced cells one unit apart, each with
sand at height $h_j$  and local gradient
$z_j=h_j-h_{j+1}$. There is a repose gradient $z_R$ 
  below which the sandpile is always stable, and with respect to which
heights $h_j$ and the gradients $z_j$ are measured.  
Each cell is assigned a critical gradient
$z_{cj}$.
 If the local gradient $z_j$ exceeds this, the sand is redistributed
to neighbouring cells and iteration produces an avalanche.
The critical gradients on each of the $N$ nodes
are selected from a  uniform (``top hat") probability distribution $P(z_{cj})$.
   $P(z_{cj})$
is generated by choosing the $z_{cj}$ at random with uniform
probability from the range $[a,b]$, and the integral of $P(z_{cj})$ over
all $z_{cj}$ is unity.

 Sand is added   at cell 1 at a rate $g$,
 and we normalize length and time
 to the mean loading rate.
  If the critical gradient is exceeded at cell 1, the
 sand is redistributed. Our redistribution rule
 is conservative and instantaneous: sand will propagate to cell 2
 and if the local critical gradient there is exceeded, to cell 3 and so on.
 Within this avalanche, the sand is instantaneously
 ``flattened" back to the angle of repose
 at which the sandpile is always stable.
Propagation of an ongoing avalanche from one cell ($k$) to the next
($k+1$) thus occurs if
\begin{equation}
h_k-h_{k+1}>z_{ck}
\end{equation}
In consequence  a quantity of sand $\Delta$ is deposited on the
next cell:
\begin{equation}
h_{k+1}^*=h_{k+1}+\Delta
\end{equation}
so that the gradient at $k$ relaxes to the angle of repose
(here normalized to zero)
\begin{equation}
h_k^*-h_{k+1}^*=z_R=0
\end{equation}
Since all cells within the ongoing avalanche $1,2...,k$ are
at the  angle of repose following this conservative redistribution
of sand to the $k+1$ cell, we require the heights of all these cells
to become:
\begin{equation}
h_{1..k}^*=h_{1..k}-\frac{\Delta}{k}
\end{equation}
The iterative procedure described above
(where superscript * denotes an intermediate step)
continues until the avalanche reaches
a cell where
the gradient is subcritical.
 The critical gradients at cells within
 the flattened post-avalanche region are then rerandomized as above. More
 sand is added then at cell
 1 until this end cell again becomes unstable, triggering another avalanche.
 An avalanche may be entirely an internal
 rearrangement of sand or may continue until it spreads across all $N$ cells.
   We call the latter a systemwide discharge, in which case the entire
 sandpile is emptied and returns to the angle of repose.

 This relaxation rule allows  propagation of
information (correlation) across the avalanche and therefore
potentially on all length scales in the sandpile. This
permits the possibility of scale free self organising behaviour in a one
dimensional system, whereas other SOC reorganisation rules
 (Bak {\em et al.}, 1987; Jensen, 1998)  have typically required two dimensions.

 The total energy dissipated by an avalanche (internal or
 systemwide) is just the difference in the potential energy
 in the entire sandpile before and after the avalanche
 \begin{equation}
 d\epsilon=\sum_{j=1}^N h_j^2 \mid _{before}-\sum_{j=1}^N h_j^2 \mid _{after}
 \label{de}
 \end{equation}

 An example of a time series for the energy is shown in Figure 1. The
5000 cell sandpile was loaded slowly ($g=0.001$) with respect to the
mean value of the $z_{cj}$. The
critical gradients  were uniformly and randomly distributed in
the range $[0.5,1.5]$. With the
angle of repose normalized to zero, the time evolution
is by systematic growth
as sand is added, interspersed with systemwide avalanches
where the energy falls back to zero, and internal avalanches
where the energy is reduced to some nonzero value.

The statistics of the energy released in internal and systemwide avalanches
for two longer runs of this sandpile under more realistic
conditions of fluctuating input are
shown in Figure 2, with the normalized
probability distribution
$P(d\epsilon)$ for both internal and systemwide  avalanches plotted  as a
single population. 
As in all
sandpile runs  here, the populations comprise
over $5\times 10^5$ internal and $2\times 10^4$ systemwide avalanches.

The two runs in figure 2 are for slow ($<g>=0.001$, diamonds) and fast ($<g>=10$,
circles) mean loading rates, giving  an indication of the expected behaviour of
a system with strong variation in the driver, such as  the
solar wind driven magnetosphere (see also (Watkins {\em et al.}, (1999)).
For both values of inflow
rate  internal avalanches show  distinct  inverse power law regions with a
turndown at small $d\epsilon$, whereas the systemwide
events which have a characteristic mean  (Chapman {\em et al.}, 1998, 1999) 
appear as a bump at the high energy end. 
The distinct behaviour of the systemwide avalanches, independent of
the inflow rate, shown here
is a necessary condition for applicability to the magnetosphere  (Chapman {\em et al.}, (1998)).
The internal avalanches show different behaviour under slow and fast loading.
 Essentially, the smaller events are   destroyed as we increase the average 
 loading rate,
making larger events more probable. Hence the normalized probability of
larger events
shows an increase on the plot. Importantly, their power law slope (here of
index $-1$) is preserved and is thus a robust feature that should be
apparent in observations under variable loading.

We  now discuss the internal avalanches in more detail.

\subsection{System scales and power law index}

Two  lines $\alpha (d\epsilon)^{-\gamma}$ and
$\beta (d\epsilon)^{-1}$ are drawn on Figure 2 (and all subsequent figures).
The values $\alpha=0.25, \gamma=0.65$ and $\beta=2$ are an  approximate
 fit through the points.
Under slow loading the  sandpile exhibits two distinct regimes.
In the case where $P(z_{cj})=\delta(z-a)$, ($a$ any positive constant) the
sandpile evolution with time  has been found analytically
and, if  the system is normalized to have total
length unity, it can be shown that $P(d\epsilon)=(d\epsilon)^{-1}$
 (Helander {\em et al.}, 1999). The arguments of Helander {\em et al.}, (1999))
lead us to expect a region of power law index $-1$   
in $P(d\epsilon)$  for  sandpile
with $P(z_{cj})$ of finite width. We might also anticipate that
as the width of  $P(z_{cj})$ is decreased more of the
total range of $P(d\epsilon)$ would be characterized by a power law index $-1$,
but surprisingly this is not so.
 
The $P(d\epsilon)$ for four sandpile runs are  superposed in Figure 3,
differing only by the choice of $P(z_{cj})$.
For three of these the same mean critical slope $<z_{cj}>=1$ but three different
widths ($0.01, 0.1, 1$) have been used. 
 The fourth
also has $P(z_{cj})$ with width $0.1$, but has a
different mean $<z_{cj}>=0.1$. In this latter case we have rescaled 
$d\epsilon \rightarrow d\epsilon\times 100$
since, on average, the heights of sand needed for instability
will be smaller by an order of magnitude, so that (5) will yield
values of $d\epsilon$ that are on average smaller by two orders of magnitude.
Figure 3  suggests that all
features of the probability distribution are robust against
the choice of $P(z_{cj})$, which effectively represents the local
condition for instability.

Avalanches dissipating smaller amounts of energy might be expected
to extend over smaller length scales, illustrated  in Figures 4-6 
where we replot the
data shown in Figure 3, showing only the
contribution from  successively longer avalanches.
Avalanches with
lengths $>1,8,64$
respectively are shown.

Independent of the details of $P(z_{cj})$ we see that the
power law index $\sim -0.65$  corresponds to avalanches that
extend over less than $\sim 64$ cells.
Figure 4 then shows that the drop
at $d\epsilon \sim 1$ in Figure 3 corresponds to avalanches
that are one cell in length.

The sandpile thus has three distinct regimes in its
statistics: single cell avalanches
that (as  expected) are not power law;
avalanches smaller than $\sim$ 64 cells, with power law index $\sim -0.65$,
which  may  reflect the discrete nature
of the grid; and avalanches longer than $\sim 64$ cells and up to the
system size, with
power law index $-1$, which may approach a continuous limit for the system.

The interval distribution (that is, the probability distribution of time intervals
between events) provides further insight into the difference between avalanches on the small and the large scale.
In figures 7-10 we show  a series of interval distributions from the single sandpile
run with constant slow fuelling $g=0.001$ and  with top hat  probability distribution
for the critical gradients $z_c=[0.95,1.05]$. The figures  show the 
time intervals between sucessively larger avalanches, that is, of
all avalanches of length $>1,4,8$ and $64$. Since sand is always added at
cell 1, and hence instability always occurs first at cell 1
a plot of the interval distribution for  all avalanches (not shown) simply
reproduces the probability distribution for $z_c$, ie, the the distribution of
times between sucessive avalanches, all of which are triggered at the first cell. 
As we selectively plot the distributions of time intervals between
longer avalanches, that have propagated further down the sandpile, we see the effect
of the interaction of more cells in the sandpile.

Figure 7 shows intervals between all avalanches that have propagated beyond cell 1
(ie length $>1$). Here we see two characteristic timescales corresponding to
avalanches that stop at cell 2, and those propagating beyond cell 2. Time is normalised
to the inflow rate (such that unit sand is added to the sandpile in unit time)
so that avalanches that reach cell 2 and stop will only occur after sufficient sand
has been added to exceed the critical gradient at cell 1, which has mean value 1.
Hence the minimum time interval in this case is $>1$.
As we increase the miminum avalanche length considered, the minimum 
time interval also increases correspondingly. The detailed behaviour
becomes complex for lengths >1 but less than 64, that is, including
avalanches which dissipate energy according to power law index $\sim -0.65$.
The general trend however is for an increasing number of characteristic time
intervals to appear as we only consider avalanches of increasing length.
When we exclude avalanches that dissipate energy according to power law index $\sim -0.65$ 
by only considering avalanches of length $>64$ (figure 10) the interval distribution
has become continuous  with cutoff at time interval
65 as we would expect. The large scale avalanches identified as those
dissipating energy with probability distribution that is  power law index $-1$ 
therefore correspond to this continuous limit. Unlike
laboratory plasmas (see  Chapman {\em et al.}, (1999))
these large scale events are expected to be relevant to astrophysical systems and
are expected to be the robust observable in the case of the magnetosphere 
which has strong, variable loading.

\section{Conclusions}
A simple one dimensional sandpile model has been developed  
with two distinct characteristics in the probability
distribution of energy discharges.
For internal reorganisation there are two distinct 
inverse power law regimes, whilst for systemwide discharges (flow of sand
out of the system) the probability distribution has a sharply-defined mean.
Our model may be applied to  magnetospheric
dynamics  (Chapman {\em et al.}, (1998)), for example in reconciling 
the apparent paradox of power law indexes
in internal dynamics with substorm event statistics which have peaked 
distributions. 

Under slow loading the internal dynamics
exhibits two  regimes which have inverse power law
statistics of
index $\sim -0.65$ and $-1$, corresponding to reconfigurations on distinct
length
scales. Short length scales may arise  
  from the discrete nature of the grid, while we also see longer scales,
up to the system size,  that effectively  approach a continuous
limit of the model. We find a transition between these regimes  at
avalanche lengths of about 64 cells.  

For space  plasma systems observations taken over long
periods are required to test for possible inverse power law statistics. The loading of
the system (in the case of the magnetosphere, the solar wind) is often
characterised by both strong variability about a mean,
and a large dynamic range of mean energy input.
The inverse power law form of the statistics of large internal 
avalanches has been shown to be
robust under fast loading. The effect of large loading
rates is to exclude events which dissipate small amounts
of energy, which in our model
results in a single  inverse power law regime with downturn
 at lower energies.
We  expect such inverse power law avalanche
distributions to be a persistent feature in long runs of data that include
``fast" inflow conditions if the underlying system is governed by SOC.

Acknowledgements: The authors thank R. O. Dendy for stimulating
discussions and Giuseppe Consolini for a preprint
of (Consolini, 1999). SCC was supported by a Particle Physics and Astronomy Research Council
lecturer fellowship.
\newcommand{\jgr}{{\it J. Geophys. Res.,}}
\newcommand{\grl}{{\it Geoph. Res. Lett.}}


\newpage
\begin{figure}
\centering
\psfig{figure=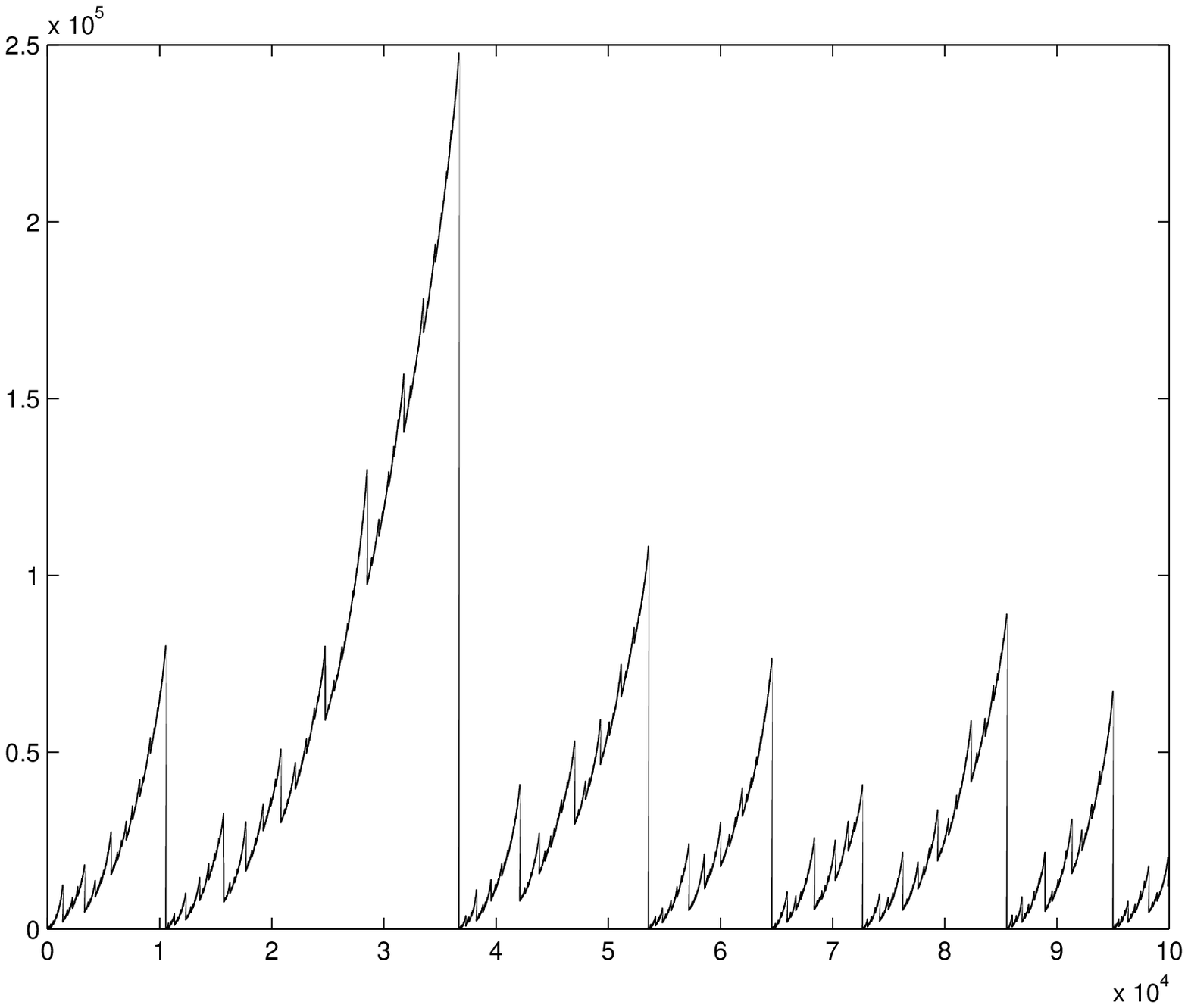,width=.9\textwidth}
\caption{The time evolution of energy in the 5000 cell sandpile, for
fuelling $g=0.001$ and probability distribution
for the critical gradients that is top hat in the range $[0.5,1.5]$.
} \end{figure}

\begin{figure}
\centering
\psfig{figure=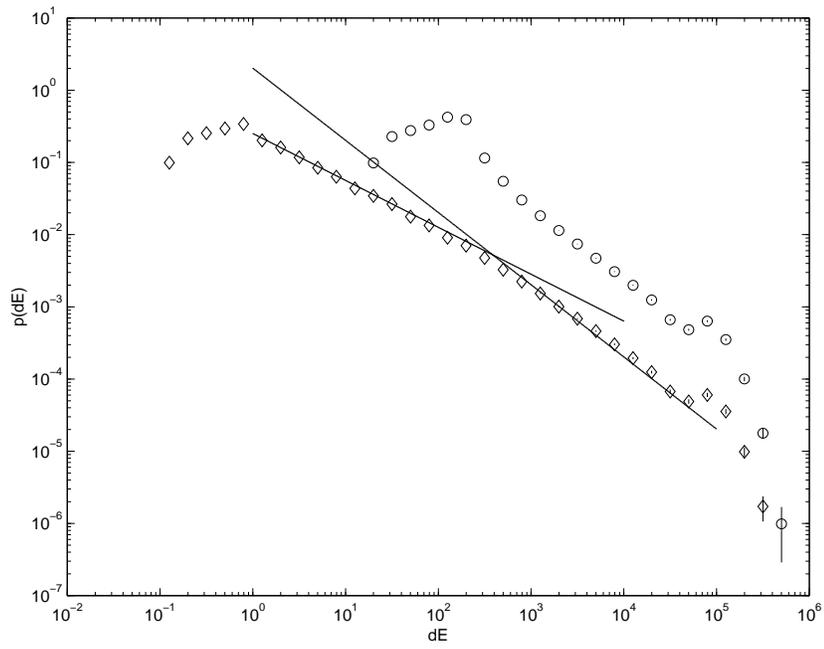,width=.9\textwidth}
\caption{The probability density of all internal
and systemwide  avalanches for  a 5000 cell sandpile with variable
fuelling $<g>=0.001$ ($\diamond$) and $<g>=10$ ($\circ$) , with  probability dis
tribution
for the critical gradients that is top hat in the range $[0.5,1.5]$.
} \end{figure}

\begin{figure}
\centering
\psfig{figure=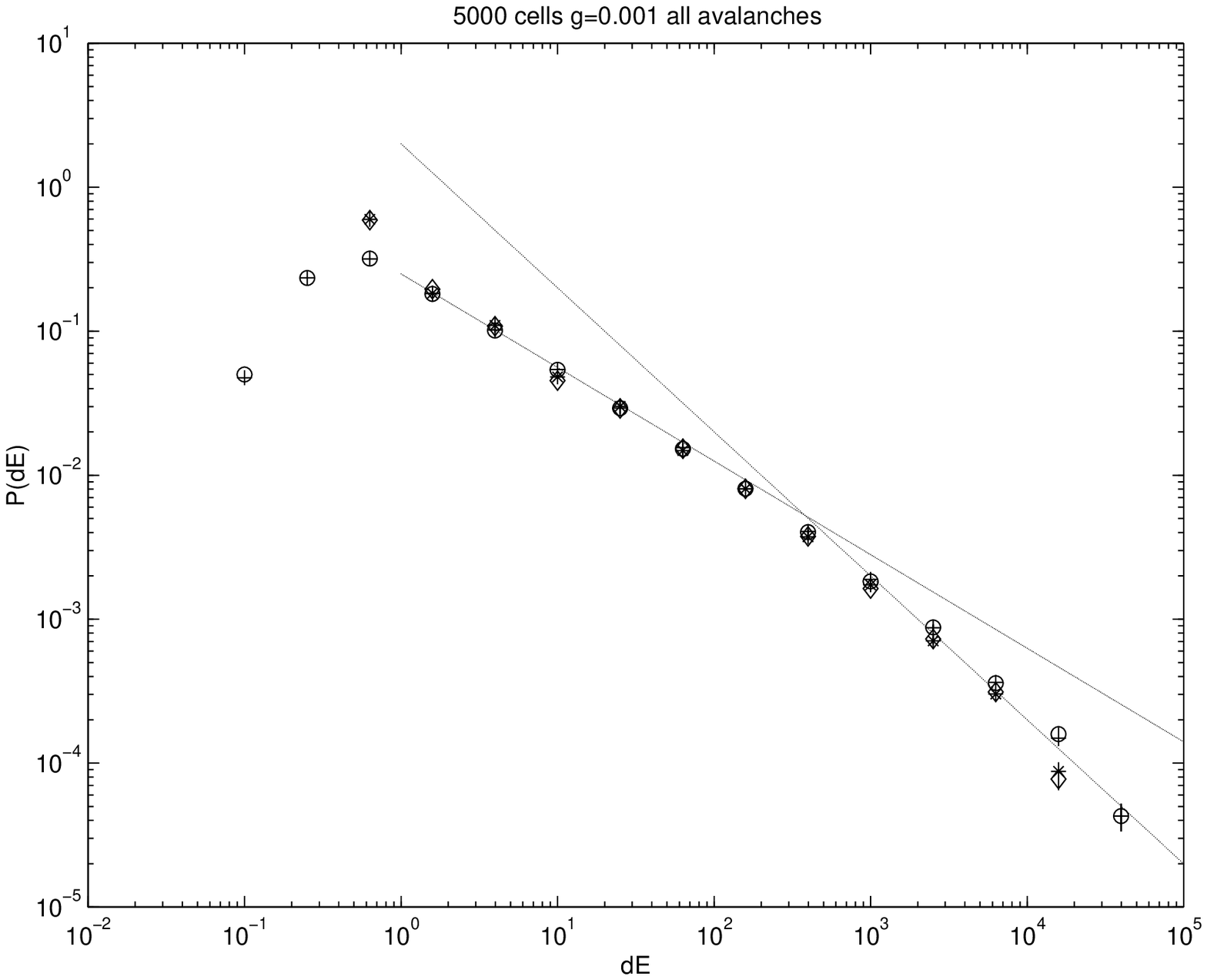,width=.9\textwidth}
\caption{The probability density of all internal
 avalanches for  a 5000 cell sandpile with constant
 fuelling $g=0.001$ and four different runs with probability distributions
 for the critical gradients that are top hat: $\circ \equiv [0.5,1.5]$,
 $\ast\equiv [0.95,1.05]$, $\diamond\equiv [0.995,1.005]$ and
 (with rescaling $d\epsilon\rightarrow d\epsilon\times 100$)
 $+\equiv [0.05,0.15]$.
 } \end{figure}

\begin{figure}
\centering
\psfig{figure=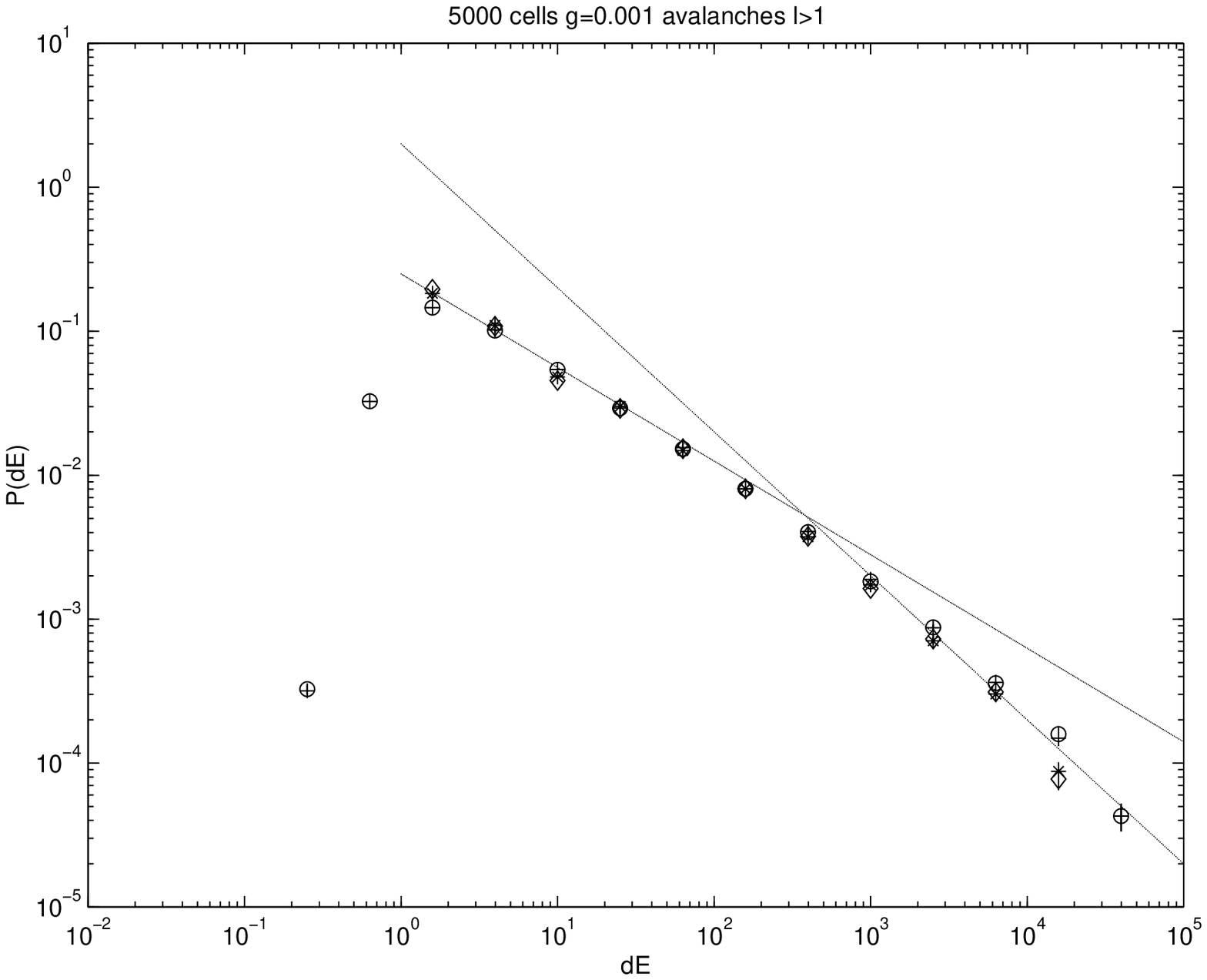,width=.9\textwidth}

\caption{The probability density of  internal
avalanches of length greater than 1 for  a 5000 cell sandpile with constant
fuelling $g=0.001$ and four different runs with probability distributions
for the critical gradients that are top hat: $\circ \equiv [0.5,1.5]$,
$\ast\equiv [0.95,1.05]$, $\diamond\equiv [0.995,1.005]$ and
(with rescaling $d\epsilon\rightarrow d\epsilon\times 100$)
$+\equiv [0.05,0.15]$.
} \end{figure}

\begin{figure}
\centering
\psfig{figure=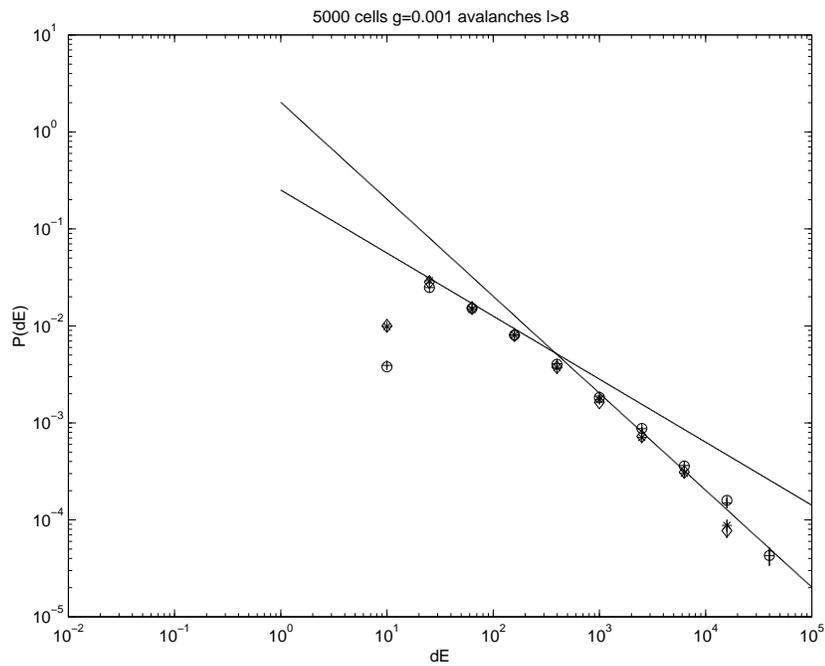,width=.9\textwidth}

\caption{The probability density of  internal
avalanches of length greater than 8 for  a 5000 cell sandpile with constant
fuelling $g=0.001$ and four different runs with probability distributions
for the critical gradients as in the previous figure.
} \end{figure}

\begin{figure}
\centering
\psfig{figure=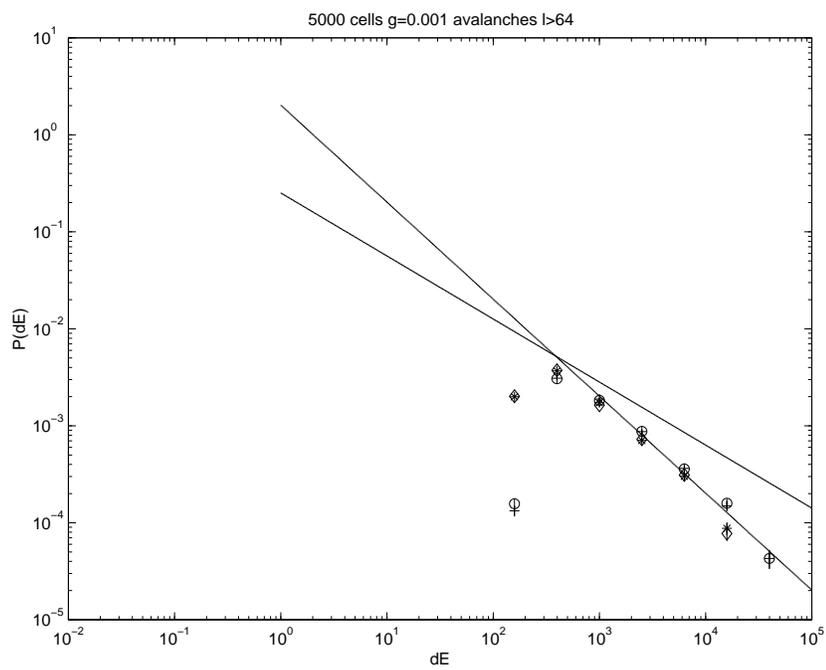,width=.9\textwidth}

\caption{The probability density of  internal
avalanches of length greater than 64 for  a 5000 cell sandpile with constant
fuelling $g=0.001$ and four different runs with probability distributions
as in the previous figure.
} \end{figure}

\begin{figure}
\centering
\psfig{figure=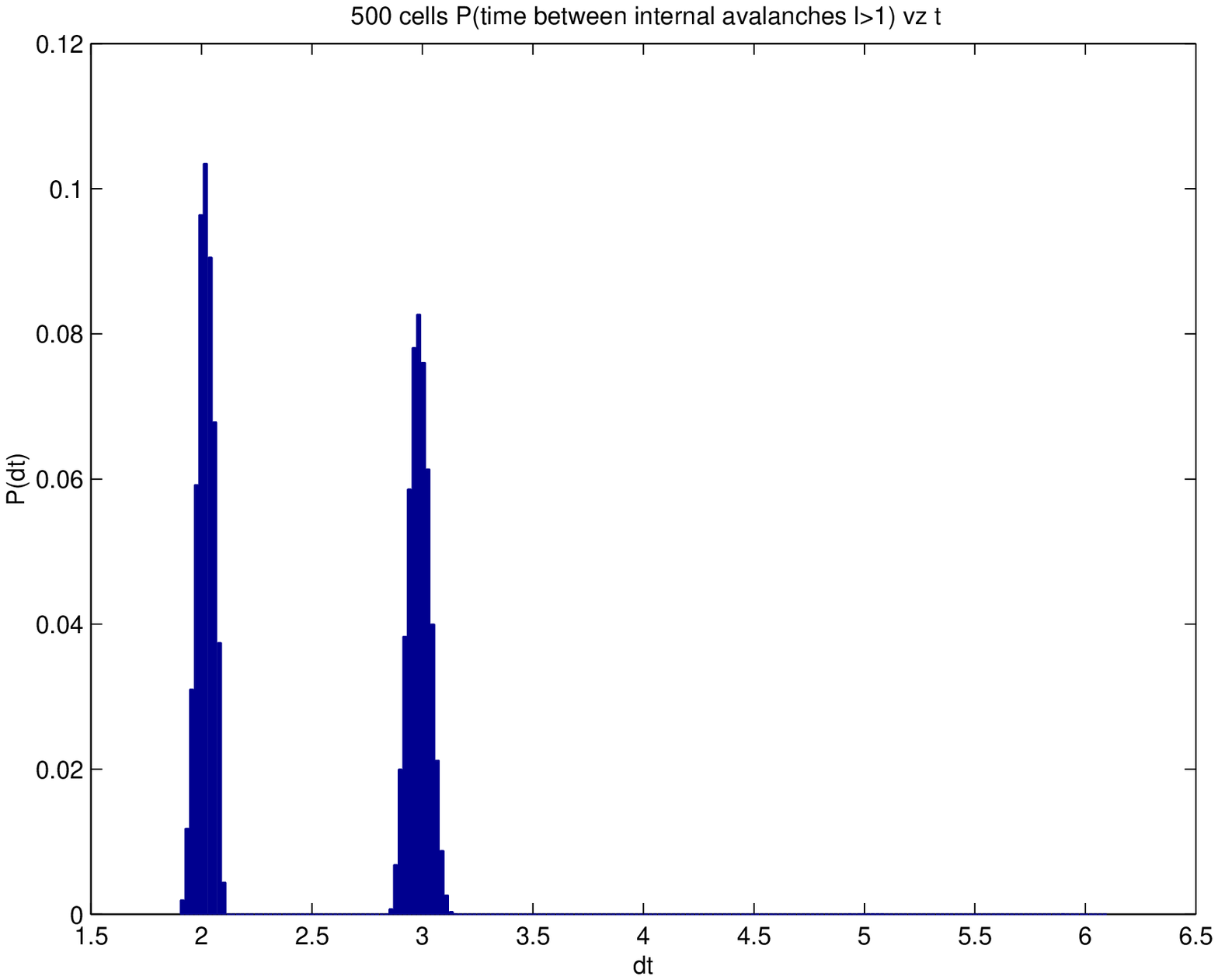,width=.9\textwidth}
\caption{The probability distribution of time intervals between avalanches
of length >1, with time normalized to the constant fuelling rate. The
fuelling $g=0.001$ and probability distribution
for the critical gradients that is top hat in the range $[0.5,1.5]$.
} \end{figure}

\begin{figure}
\centering
\psfig{figure=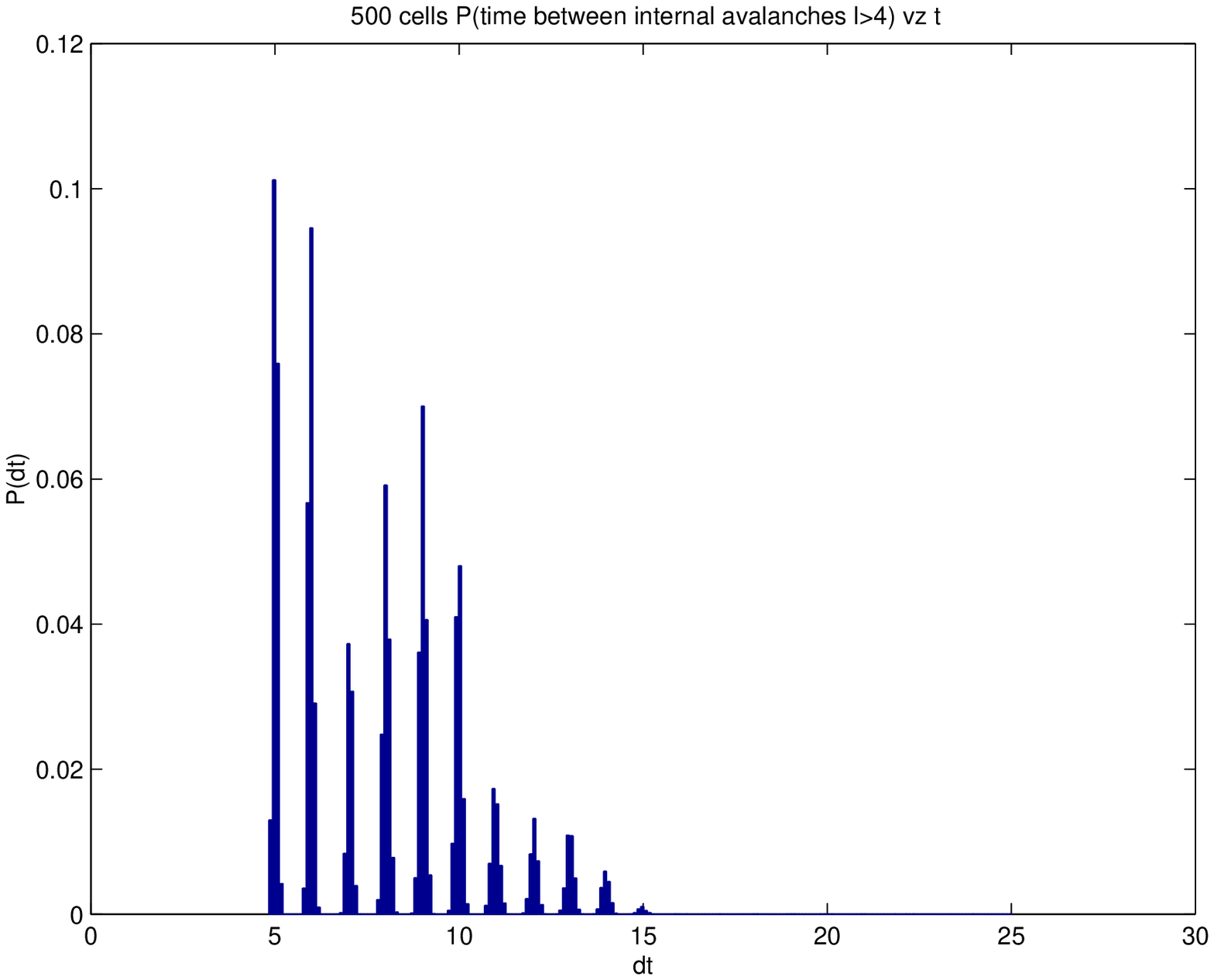,width=.9\textwidth}
\caption{The probability distribution of time intervals between avalanches
of length >4, with time normalized to the constant fuelling rate. The
fuelling $g=0.001$ and probability distribution
for the critical gradients that is top hat in the range $[0.5,1.5]$.
} \end{figure}

\begin{figure}
\centering
\psfig{figure=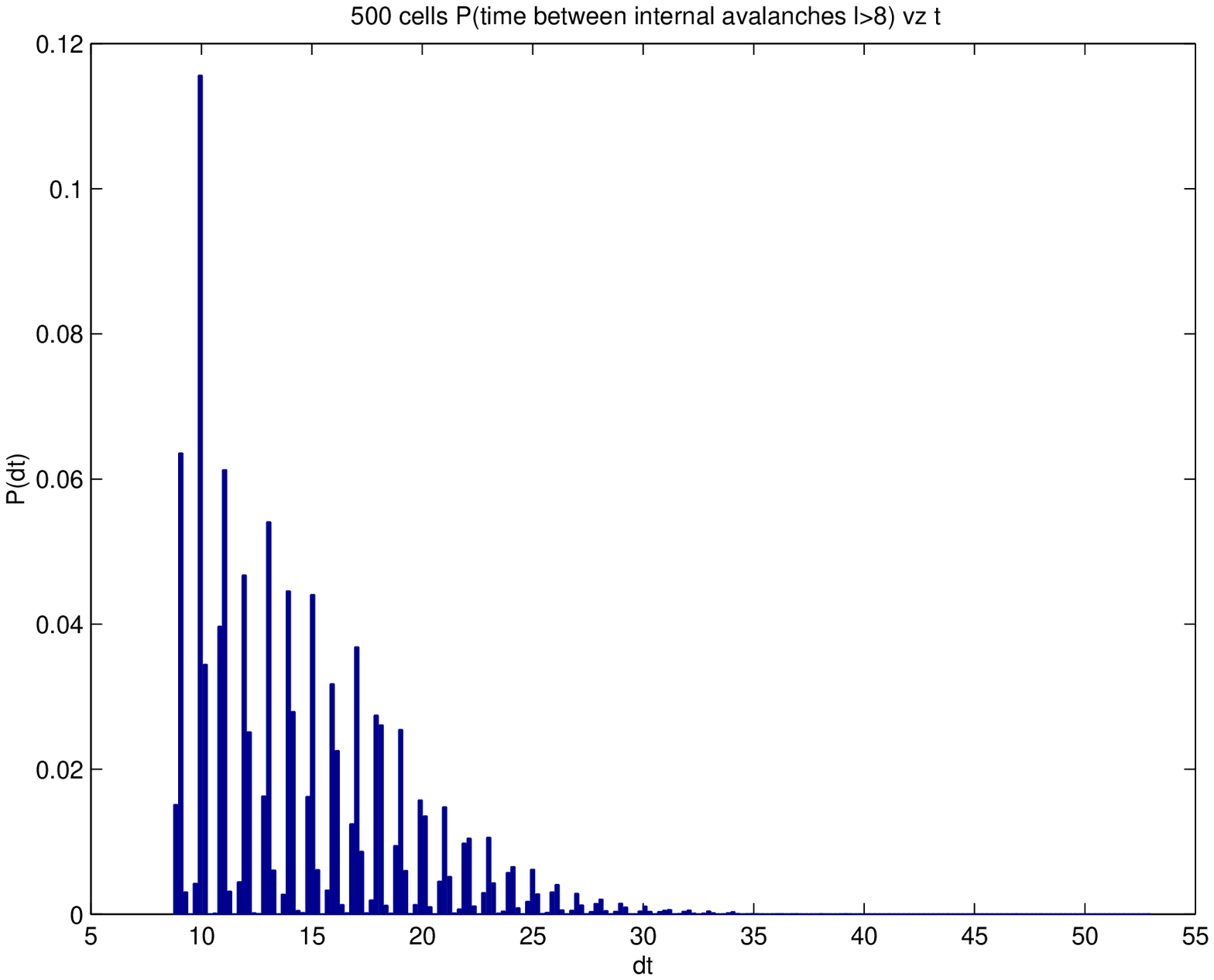,width=.9\textwidth}
\caption{The probability distribution of time intervals between avalanches
of length >8, with time normalized to the constant fuelling rate. The
fuelling $g=0.001$ and probability distribution
for the critical gradients that is top hat in the range $[0.5,1.5]$.
} \end{figure}
\begin{figure}
\centering
\psfig{figure=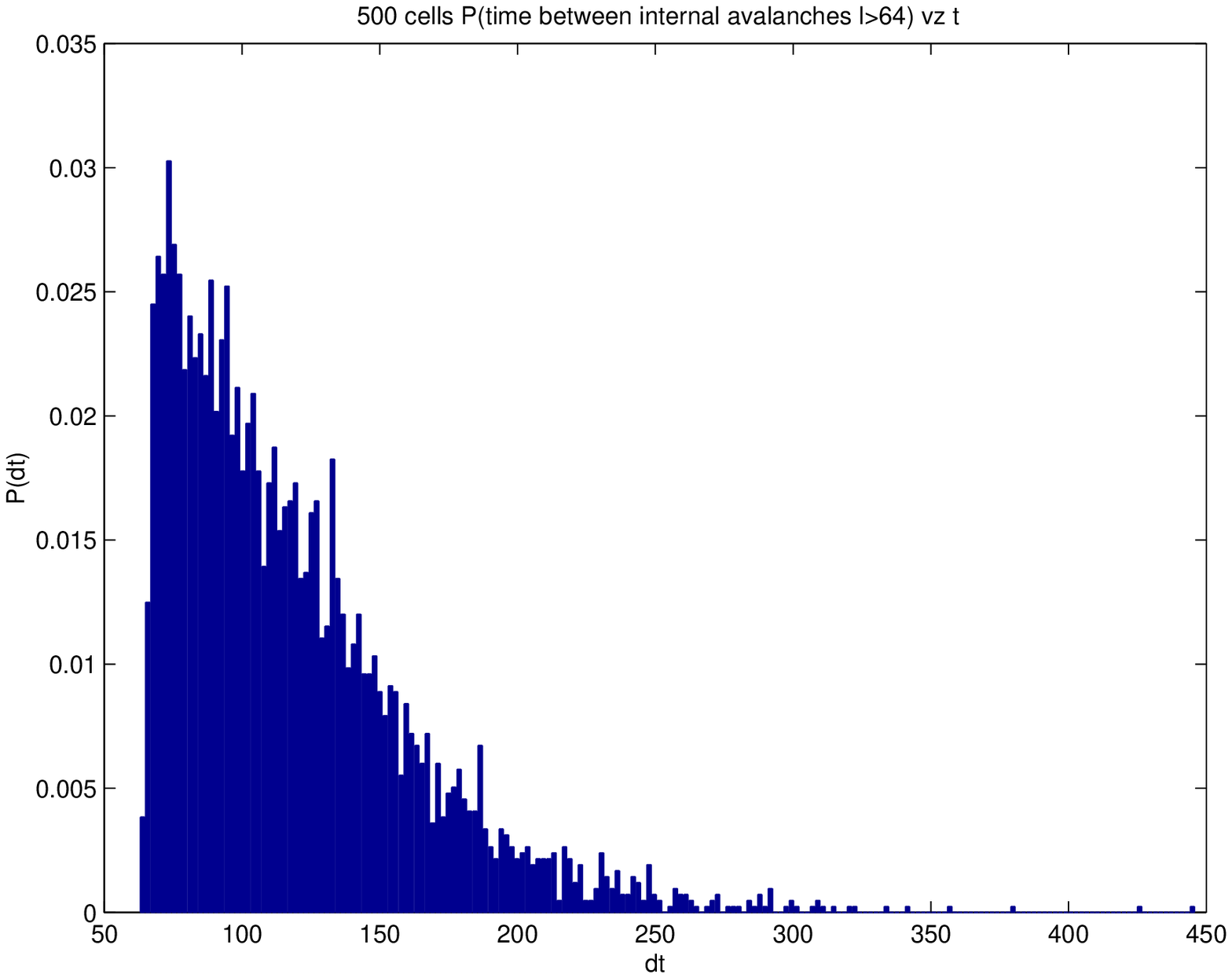,width=.9\textwidth}
\caption{The probability distribution of time intervals between avalanches
of length >64, with time normalized to the constant fuelling rate. The
fuelling $g=0.001$ and probability distribution
for the critical gradients that is top hat in the range $[0.5,1.5]$.
} \end{figure}


\begin{thebibliography}{99}
\bibitem{angel} Angelopoulos V., Coroniti F. V.,  Kennel C. F. , Kivelson M. G.,
Walker R. J., Russell C. T., McPherron R. L.,  Sanchez E.,  Meng C. I.,
Baumjohann W., Reeves G. D.,  Belian R. D., Sato N., Friis-Christensen E.,
 Sutcliffe P. R., Yumoto K. and  Harris T. (1996) Multipoint
 analysis of a bursty bulk flow event on April 11, 1985. {\em Journal of Geophysical
 Research} {\bf 101}, 4967-4989.

\bibitem{btw87}
 Bak P.,  Tang C. and Weisenfeld K. (1987)
Self-organised criticality: An explanation of 1/f noise.
 {\em Physical  Review  Letters} {\bf 59}, 381--384.

\bibitem{Bak:Tang}
Bak P., Tang C. and Wiesenfeld, K. (1988) Self-organised criticality.
{\em Physical Review} {\bf A38}, 364--374. 

\bibitem{chang92}
 Chang T.  S. (1992)
 Low dimensional behaviour and symmetry breaking of
 stochastic systems near criticality - can these effects be observed in 
 space and in the laboratory ?. {\em IEEE Transactions on  Plasma Science}, {\it 20}, 691--694.

\bibitem{chang98a}
 Chang T.  S. (1998a) Multiscale intermittent turbulence in the magnetotail.
 In {\em Substorms-4.} ed.  S. Kokubun   and Y.  Kamide, pp 431--436. Terra Scientific
Publishing Company/Kluwer Academic Publishers, Tokyo.

\bibitem{chang98b}
 Chang T.  S. (1998b) Sporadic localised reconnections and multiscale 
 intermittent turbulence in the magnetotail. In {\em
 Geospace Mass and Energy Flow: Results from the International Solar
 Terrestrial Physics Program.} ed. J. L. Horwitz, D. L. Gallagher
 and W. K. Peterson, pp. 193-199. Geophysical Monograph 104,
American Geophysical Union, Washington, D.C.

\bibitem{chapman98}
 Chapman S. C., Watkins N. W.,   Dendy R. O., Helander P. and Rowlands G. (1998)
A simple avalanche model as an analogue
for magnetospheric activity. {\em Geophysical Research Letters} {\bf 25}, 2397--2400.

 
\bibitem{chapman99}
 Chapman S. C.,     Dendy R. O.  and Rowlands G. (1999) A sandpile model with
 dual scaling regimes for laboratory, space and astrophysical plasmas. 
  {\em Physics of Plasmas}, {\bf 6}, 4169 .  

\bibitem{Consolini97}
 Consolini G. (1997) Sandpile cellular automata and magnetospheric dynamics.
 In {\em Proceedings volume  58, ``Cosmic Physics in the Year 2000"}. eds. S.  Aiello, N.  Iucci,
G.  Sironi, A.  Treves and U.  Villante, pp 123--126. Societa Italiana di
Fisica, Bologna, Italy.

\bibitem{consolini99}
Consolini G. (1999) Avalanches, scaling and 1/f noise in magnetospheric dynamics. {\em
Physical Review Letters}, submitted.


 
\bibitem{dendy97} Dendy R. O.  and Helander P. (1997)
Sandpiles, silos and tokamak phenomenology: a brief review.
 {\em Plasma Physics and Controlled Fusion}, {\bf 39},
1947--1961.


\bibitem{Dendy:Tagger}
Dendy R.O., Helander P. and Tagger M. (1998)
On the role of self-organised criticality in accretion systems.
{\it Astronomy and  Astrophysics} {\bf 337}, 962--965.

\bibitem{freeman99}
Freeman M. P., Watkins N. W. and Riley. D. J. (1999) Evidence for a solar wind origin of the power
law burst lifetime distribution of the AE indices. {\em Geophysical Research Letters}, submitted.
 
\bibitem{helander99}
 Helander P., Chapman S. C., Dendy R. O., Rowlands G.
and  Watkins, N. W. (1999) Exactly solvable sandpile with fractal
avalanching. {\em Physical Review}, {\bf E59}, 6356--6360.  

\bibitem{hoshino}
Hoshino M., Nishida A., Yamamoto T. and Kokubun S. (1994)
Turbulent magnetic field in the distant magnetotail -- bottom up process of plasmoid
formation ?
  {\em Geophysical Research Letters}, {\bf 21}, 2935.

\bibitem{jensen98}
 Jensen H.  J. (1998) {\em
Self-Organised Criticality:  Emergent Complex Behaviour in Physical
and Biological Systems,} Cambridge University Press, Cambridge.

\bibitem{kad} Kadanoff L. P., Nagel S. R., Wu L. and Zhou, S. (1989)
Scaling and universality in avalanches.
 {\it Physical Review },
{\bf A39}, 6524
 
\bibitem{lue95}  Lu E. (1995) Avalanches in continuum dissipative 
systems.
 {\it Physical Review Letters}
{\bf 74}, 2511--2514

 
 
 
\bibitem{pinho1998}  Pinho S. T. R, and
 Andrade R. F. S. (1998) 
 An Abelian model for rainfall.
 {\em Physica}, {\bf A255}, 483--495

 
 

\bibitem{tsurutani90}
Tsurutani B., Sugiura M., Iyemori T., Goldstein B. E., 
Gonzalez W. D., Akosofu S.-I. and E.  J.  Smith (1990) The nonlinear response of
AE to the IMF $B_s$ driver:  a spectral break at 5 hours.
{\em Geophysical Research Letters}, {\bf 17}, 279--282.
 


 
\bibitem{nickgrl} Watkins, N. W.,  Chapman S. C., Dendy R. O.  
and  Rowlands G. (1999) Robustness of collective behaviour
in a strongly driven avalanche model: magnetospheric 
implications. {\it Geophysical Research Letters},  {\bf 26}, 2617-2620.


 


\end{thebibliography}
\end{document}